\documentclass[useAMS,usenatbib]{mn2e}
\usepackage{graphicx}
\usepackage{float}
\title[Long-term evolution and X-ray outburst of SGR 0501+4516]{Long-term evolution, X-ray outburst and optical/infrared emission of SGR 0501+4516}
\author[O. Benli, \c{S}. \c{C}al{\i}\c{s}kan and \"{U}. Ertan]{O. Benli\thanks{E-mail: onurbenli@sabanciuniv.edu}, \c{S}. \c{C}al{\i}\c{s}kan and \"{U}. Ertan \\
Sabanc\i\ University, 34956, Orhanl\i\, Tuzla, \.Istanbul, Turkey}
\begin{document}

\date{2015 January 11}


\maketitle

\newcommand {\gtrsim} {\ {\raise-.5ex\hbox{$\buildrel>\over\sim$}}\ }
\newcommand {\lesssim} {\ {\raise-.5ex\hbox{$\buildrel<\over\sim$}}\ }

\def\la{\raise.5ex\hbox{$<$}\kern-.8em\lower 1mm\hbox{$\sim$}}
\def\ga{\raise.5ex\hbox{$>$}\kern-.8em\lower 1mm\hbox{$\sim$}}
\def\be{\begin{equation}}
\def\ee{\end{equation}}
\def\ba{\begin{eqnarray}}
\def\ea{\end{eqnarray}}
\def\m{\mathrm}
\def\d{\partial}
\def\R{\right}
\def\L{\left}
\def\a{\alpha}
\def\acold{\alpha_\mathrm{cold}}
\def\ahot{\alpha_\mathrm{hot}}
\def\Mdotstar{\dot{M}_\ast}
\def\Omegastar{\Omega_\ast}
\def\OmegaK{$\Omega$_{\mathrm{K}}}
\def\Mdotin{\dot{M}_{\mathrm{in}}}
\def\Mdot{\dot{M}}
\def\Edot{\dot{E}}
\def\Pdot{\dot{P}}
\def\Msun{M_{\odot}}
\def\Lin{L_{\mathrm{in}}}
\def\Lcool{L_{\mathrm{cool}}}
\def\Lacc{L_{\mathrm{acc}}}
\def\Ldiss{L_{\mathrm{diss}}}
\def\Rin{R_{\mathrm{in}}}
\def\rin{r_{\mathrm{in}}}
\def\rlc{r_{\mathrm{LC}}}
\def\rout{r_{\mathrm{out}}}
\def\rco{r_{\mathrm{co}}}
\def\Rout{R_{\mathrm{out}}}
\def\Ldisc{L_{\mathrm{disc}}}
\def\Lx{L_{\mathrm{x}}}
\def\Md{M_{\mathrm{d}}}
\def\cs{c_{\mathrm{s}}}
\def\dEb{\delta E_{\mathrm{burst}}}
\def\dEx{\delta E_{\mathrm{x}}}
\def\Bstar{B_\ast}
\def\Bb{\beta_{\mathrm{b}}}
\def\Be{\beta_{\mathrm{e}}}
\def\Rc{\R_{\mathrm{c}}}
\def\rA{r_{\mathrm{A}}}
\def\rp{r_{\mathrm{p}}}
\def\Tp{T_{\mathrm{p}}}
\def\dMin{\delta M_{\mathrm{in}}}
\def\dM*{\delta M_*}
\def\Teff{T_{\mathrm{eff}}}
\def\Tirr{T_{\mathrm{irr}}}
\def\Firr{F_{\mathrm{irr}}}
\def\Tcrit{T_{\mathrm{crit}}}
\def\P0min{P_{0,{\mathrm{min}}}}
\def\Av{A_{\mathrm{V}}}
\def\ah{\alpha_{\mathrm{hot}}}
\def\ac{\alpha_{\mathrm{cold}}}
\def\tc{\tau_{\mathrm{c}}}
\def\p{\propto}
\def\m{\mathrm}
\def\fast{\omega_{\ast}}
\def\Alfven{Alfv$\acute{e}$n}
\def\418{SGR 0418+5729}
\def\142{AXP 0142+61}
\def\Caliskan{\c{C}al{\i}\c{s}kan}
\def\0501{SGR 0501+4516}
\def\ql{\textquotedblleft}
\def\qr{\textquotedblright}
\def\rzero{r_{\mathrm{0}}} 
\def\Bzero{B_{\mathrm{0}}} 
\def\Smax{\Sigma_{\mathrm{max}} }
\def\Szero{\Sigma_{\mathrm{0}} } 
\def\dr{$\Delta$r }
\label{firstpage}

\begin{abstract}
We have analyzed the long-term evolution and the X-ray outburst light curve of \0501~in the frame of the fallback disc model. We have shown that the X-ray luminosity, period and period derivative of this typical soft gamma repeater can be achieved by a neutron star with a large range of initial disc masses provided that the source has a magnetic dipole field of $\sim 1.4 \times 10^{12}$ G on the pole of the star. At present, the star is accreting matter from the disc, which has an age $\sim 3 \times 10^4$ yr, and will remain in the accretion phase until $t \sim 2$--$5 \times 10^5$ yr depending on the initial disc mass. With its current rotational rate, this source is not expected to give pulsed radio emission even if the accretion on to the star is hindered by some mechanism. The X-ray enhancement light curve of \0501~can be accounted for by the same model applied earlier to the X-ray enhancement light curves of other anomalous X-ray pulsars/soft gamma repeaters with the same basic disc parameters. We have further shown that the optical/IR data of \0501~is in good agreement with the emission from an irradiated fallback disc with the properties consistent with our long-term evolution model.
\end{abstract}


\section{INTRODUCTION}

Anomalous X-ray pulsars (AXPs) and soft gamma repeaters (SGRs) constitute a young  neutron star population with some extreme properties. They show short ($\lesssim 1$ s) super-Eddington soft gamma burst ($\gtrsim 10^{40}$ erg s$^{-1}$). Three giant bursts ($\gtrsim 10^{44}$ erg s$^{-1}$) were observed from three different SGRs. For most AXP/SGRs, quiescent X-ray luminosities ($\Lx \sim 10^{33}$--$10^{36}$ erg s$^{-1}$) are much higher than the rotational powers $\dot{E} = I \Omega \dot{\Omega}$ of the sources. All known $\sim 20$ AXP/SGRs have the rotational periods clustered in the $2$-$12$ s range (see Mereghetti 2008 for a review of AXP/SGRs).

Some of these sources exhibit X-ray outbursts (enhancements) after the burst episodes. These X-ray outbursts can be explained in the frames of both the magnetar model (Thompson \& Duncan 1995) and the fallback disc model (Chatterjee, Hernquist \& Narayan 2000; Alpar 2001). In the magnetar model, part of the energy powering the burst is injected into the crust; the resultant heating and subsequent cooling of the crust produce the observed outburst light curve (see, e.g., Camero et al. 2014). In the fallback disc model, part of the burst energy is absorbed by the inner disc. The inner disc matter is pushed back by the burst, and piles up at a larger radius forming a density gradient. Starting from this initial condition, the disc evolves with an abruptly enhanced mass-flow and accretion rate leading to the onset of the X-ray outburst. Subsequently, the accretion rate decreases with the relaxation of the disc, and produces the decay phase of the X-ray light curve. This model can explain the the X-ray outburst light curves of different AXP/SGRs with the same basic disc parameters (\Caliskan~\& Ertan 2012).

The explanation of the long-term evolution of AXP/SGRs is also rather different in the magnetar and the fallback disc models. In the magnetar model, these sources are evolving in vacuum and are slowed down by magnetic dipole torques. With this assumption, the strength of the dipole field on the surface of the neutron star is estimated from the observed period and period derivative of the source using $B \simeq 3.2 \times  10^{19} \sqrt{P \Pdot}$ G, which gives $B > 10^{14}$ G for most AXP/SGRs. In this model, the explanation of the X-ray luminosity and the rotational properties of AXP/SGRs requires the decay of the magnetic dipole field. The required decay is very rapid for some sources (Turolla et al. 2011; Vigano et al. 2013), while it is negligible for some others (e.g. Camero et al. 2014).  The decay rate of the dipole field depends mainly on the strength of the initial crustal toroidal field. For instance, for a neutron star to reach the properties of the so-called low-B magnetar SGR $0418$+$5729$, it should have an initial toroidal field stronger than $10^{16}$ G (Turolla et al. 2011). For a second low-B magnetar Swift J1822.3--1606, the  model needs an initial crustal toroidal field with a strength of a few $10^{14}$ G.

In the fallback disc model, explanation of the optical emission and the long-term evolution of AXP/SGRs requires conventional dipole fields of young neutron stars ($10^{12}$--$10^{13}$ G). The SGR bursts are likely to be powered by strong magnetic fields ($B > 10^{14}$ G) on the surface of the star as described in the magnetar model. Nevertheless, these magnetar fields could be in the small-scale quadrupole fields rather than the dipole component as indicated by the observations of low-B magnetars. In the fallback disk model, the dipole component of the magnetic field interacts with the inner disc, while the higher multipoles do not affect the rotational evolution of the neutron star. This means that small-scale magnetar fields close to the surface of the star is compatible with the disc model. However, a hybrid model with a fallback disc around a magnetar dipole field cannot produce observed rotational properties, in particular the period clustering of AXP/SGRs (Alpar 2001; Ek{\c s}i \& Alpar 2003; Ertan et al. 2007, 2009).

In the fallback disc model, the torque acting on the neutron star is provided by the disc. The sources enter a long-lasting accretion phase at some early phase of evolution depending on the initial conditions: dipole field strength, initial period and the disc mass. The sources experience the most efficient torque in the accretion phase. The mass accretion on to the star powers the X-ray luminosity. Before the onset and after the termination of the accretion phase, the X-ray luminosity is produced by the intrinsic cooling of the neutron star. This model can produce the basic X-ray and the rotational properties of AXP/SGRs (Ertan \& Erkut 2008; Ertan et al. 2009), dim isolated neutron stars (Ertan et al. 2014), low-B magnetars (Alpar, Ertan \& \c{C}al{\i}\c{s}kan 2011; Benli et al. 2013), and the so-called \ql high-B radio pulsar\qr~PSR J1734--3333, which has an anomalous breaking index $n\simeq 1$ (\Caliskan~et al. 2013). In this model,  the dipole field strength required for all these different sources is less than $10^{13}$ G, and the low-B magnetars with apparently extreme properties are just ordinary AXP/SGRs in a relatively late phase of their long-term evolution. Consistent with these results, the typical high-energy ($> 10$ keV) spectra of AXP/SGRs can be accounted for by the bulk-motion Comptonization in the accretion columns of neutron stars with the accretion rates indicated by the X-ray luminosities of these sources (Tr\"umper et al. 2010, 2013; Kylafis, Tr{\"u}mper \& Ertan 2014).    

Recently Camero et al. (2014) performed model fits to the X-ray outburst light curve of \0501, a typical SGR, and also present the long-term evolutionary scenario expected for this source in the magnetar model. The model requires an initial dipole field of strength $\sim 3 \times 10^{14}$ G on the pole. Since the current dipole field on the pole ($\sim 4 \times 10^{14}$ G) inferred from the dipole formula is similar to the initial field, there is either negligible or no field decay for this source in this model.   

In the present work, we analyze both the enhancement light curve and the long-term evolution of \0501 with the fallback disc model. We also compare the estimated optical/infrared (IR) flux of the disc with the observed data. We describe the model parameters briefly, and give the results for the short and the long-term evolution models in Sections 2 and 3 respectively. Properties of \0501 indicated by our results are discussed and compared  to those of the low-B magnetars and \ql high-B radio pulsar\qr~ PSR J1734--3333. There is a summary of our conclusions in Section 4.

\section{Long-Term Evolution of \0501}

\0501 was discovered by {\it Swift--BAT} during a burst episode starting from 2008 August 22 (Holland et al. 2008, Barthelmy et al. 2008). The source was observed by RXTE following the Swift detection and a period of 5.762067(2) s was found from the coherent X-ray pulsations (G\"{o}\u{g}\"{u}\c{s}, Woods \& Kouveliotou 2008). The source  also showed optical pulsations with a period of 5.7622 $\pm$ 0.0003 s  which is in good agreement with the X-ray spin period (Dhillon et al. 2011). The subsequent observations with combined {\it RXTE/PCA, Swift/XRT, CXO/ACIS-S} and {\it XMM-Newton/EPIC-PN} observations revealed a period derivative of $\Pdot \simeq 5.8 \times 10^{-12}$ s s$^{-1}$ (G\"{o}\u{g}\"{u}\c{s} et al. 2010).

The minimum distance to \0501 is estimated to be $\sim$ 1.5 kpc based on a likely association of the source with the supernova remnant HB9 (Aptekar et al. 2009). We have converted the observed X-ray flux into the X-ray luminosity assuming that the source lies at a distance in the $1.5$--$5$ kpc range. A blackbody plus a power-law model fitted well to the quiescent soft X-ray spectrum of \0501~(Camero et al. 2014). The best-fitting parameters are kT = 0.52 $\pm$ 0.02 keV with a blackbody radius of 0.39 $\pm$ 0.05 km, $N_\mathrm{H} = 0.85(3) \times 10^{22}$ cm$^{-2}$ and the power law index $\Gamma = 3.84 \pm 0.06$. The bolometric luminosities obtained from these models are $4.7 \times 10^{33}$ erg s$^{-1}$ and $5.2 \times 10^{34}$ erg s$^{-1}$ for distances 1.5 and 5 kpc.  

In this section, we investigate the long-term X-ray luminosity and rotational evolution of \0501~with a fallback disc (for details and applications of this model see Ertan et al. 2009; Alpar et al. 2011; Benli et al. 2013; \Caliskan~et al. 2013). We solve the diffusion equation for a geometrically thin disc. The disc evolves interacting with the dipole field of the neutron star. When the inner disc can penetrate the light cylinder, we take the inner disc radius, $\rin$, to be equal to the \Alfven~radius, $\rA \cong (G M)^{-1/7}~\mu^{4/7} \Mdotin^{-2/7}$ where $\Mdotin$ is the rate of mass-flow to the inner disc, $\mu$ is the magnetic dipole moment and $M$ is the mass of the neutron star. When $\rA$ is found to be greater than the light-cylinder radius, $\rlc$, we set $\rin = \rlc$. The disc evolves under the effect of the X-ray irradiation. At a given time during the evolution, the disc is viscously active from $\rin$ to the radius $\rout$ at which the effective temperature is currently equal to the minimum critical temperature $\Tp$. During the long-term evolution,  $\rout$ propagates inward with decreasing X-ray irradiation flux that is defined through $\Firr = C \Mdot c^2 / (4 \mathrm{\pi} r^2$) (Shakura \& Sunyaev 1973) where $c$ is the speed of light, $\Mdot$ is the accretion rate on to the neutron star and $C$ is the X-ray irradiation efficiency parameter. The X-ray and IR analysis of AXP/SGRs in quiescence constrains $C$ into the $1$--$7 \times 10^{-4}$ range (Ertan \& \Caliskan~ 2006). In our earlier work, we obtain reasonable model curves with the minimum temperature of the active disc $\Tp \sim 100$ K. 

We use the torque model described in Ertan \& Erkut (2008). The initial parameters of the model are the dipole field strength on the pole of the star, $B_0$, the disc mass, $M_\mathrm{d}$, and the initial period $P_0$.

\begin{figure}
\includegraphics[width=.7\textwidth,angle=270]{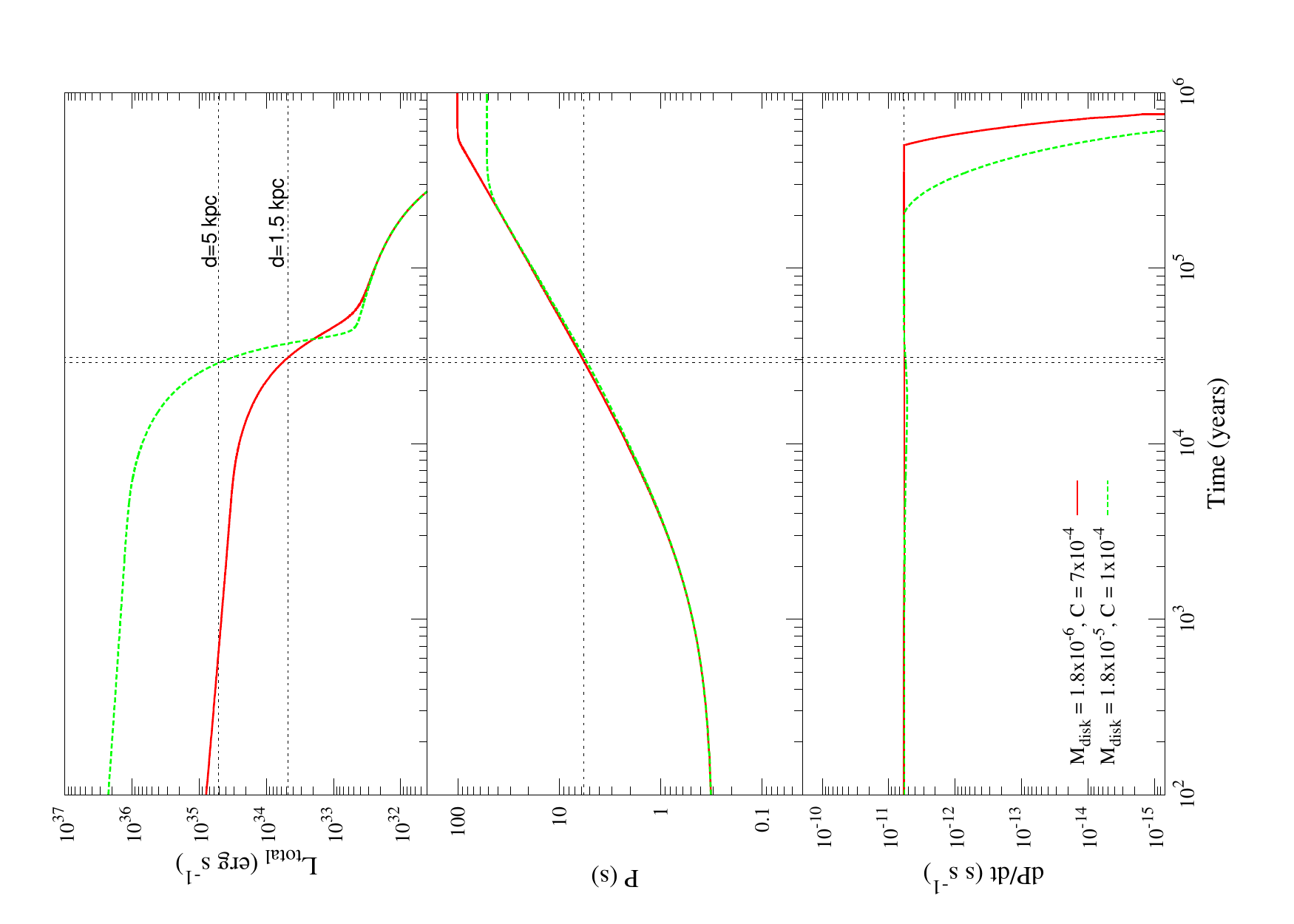}
\caption{Illustrative model curves that can represent the long-term evolution of \0501. The model sources are still accreting at present. The dipole field strength $B_0\simeq1.4 \times 10^{12}$ G for both models. The disc masses in solar mass and the values of the irradiation parameter $C$ are given in the bottom panel. Horizontal dotted lines show the properties of \0501. The lower and upper limits in the luminosity correspond to distances of $\sim1.5$ and $5$ kpc.}
\end{figure}    

The model curves given in Fig. 1 are obtained for distances 1.5 and 5 kpc with the disc masses $1.8 \times 10^{-6}$ and $1.8 \times 10^{-5} \Msun$ respectively. Independent of the disc mass, the dipole field strength on the pole of the star is well constrained to a narrow range with $B_0 \simeq 1.4 \times 10^{12}$ G which is more than two orders of magnitude weaker than that inferred from the dipole torque formula. Since the source does not come close to the rotational equilibrium, $\Pdot \propto B_0^2$~ and remains almost constant in the accretion phase (see e.g. \Caliskan~et al. 2013 for the $\Pdot$ behaviour in different phases of evolution). It is seen in Fig. 1 that the X-ray luminosity (upper panel), the period (middle panel) and the period derivative (bottom panel) are reached simultaneously by the model sources in the accretion phase at an age $\sim 3 \times 10^4$ yr which is about 2 times greater than the characteristic age ($P/2\Pdot$) of \0501. The source could remain in the accretion phase (constant $\Pdot$ phase) until an age $\sim 2$--$5 \times 10^5$ yr depending on its actual initial disc mass. The accretion epoch terminates when the inner disc cannot penetrate the light cylinder, and the system enters the tracking phase with $\rin = \rlc$. During the tracking phase, $\Pdot$ decreases with decreasing $\Mdotin$, that is, the maximum $\Pdot$ is obtained in the accretion phase. It is also possible that some of the sources could start their evolution in the tracking phase with an inefficient torque. Whether these sources can enter the accretion phase at a later time of evolution depends on the initial period, $P_0$, for a given dipole field and disk mass. If $P_0$ is below a minimum critical value the disk torque cannot sufficiently slow-down the neutron star such that the inner disk can never penetrate the light cylinder in the active lifetime of the disk. In this case, $\Pdot$ decreases continuously converging to the $\Pdot$ level of the dipole torque. These sources are likely to become radio pulsars following an evolutionary track rather different from those of accreting sources (see Fig. 2). 

\begin{figure}
\includegraphics[height=.37\textheight,angle=270]{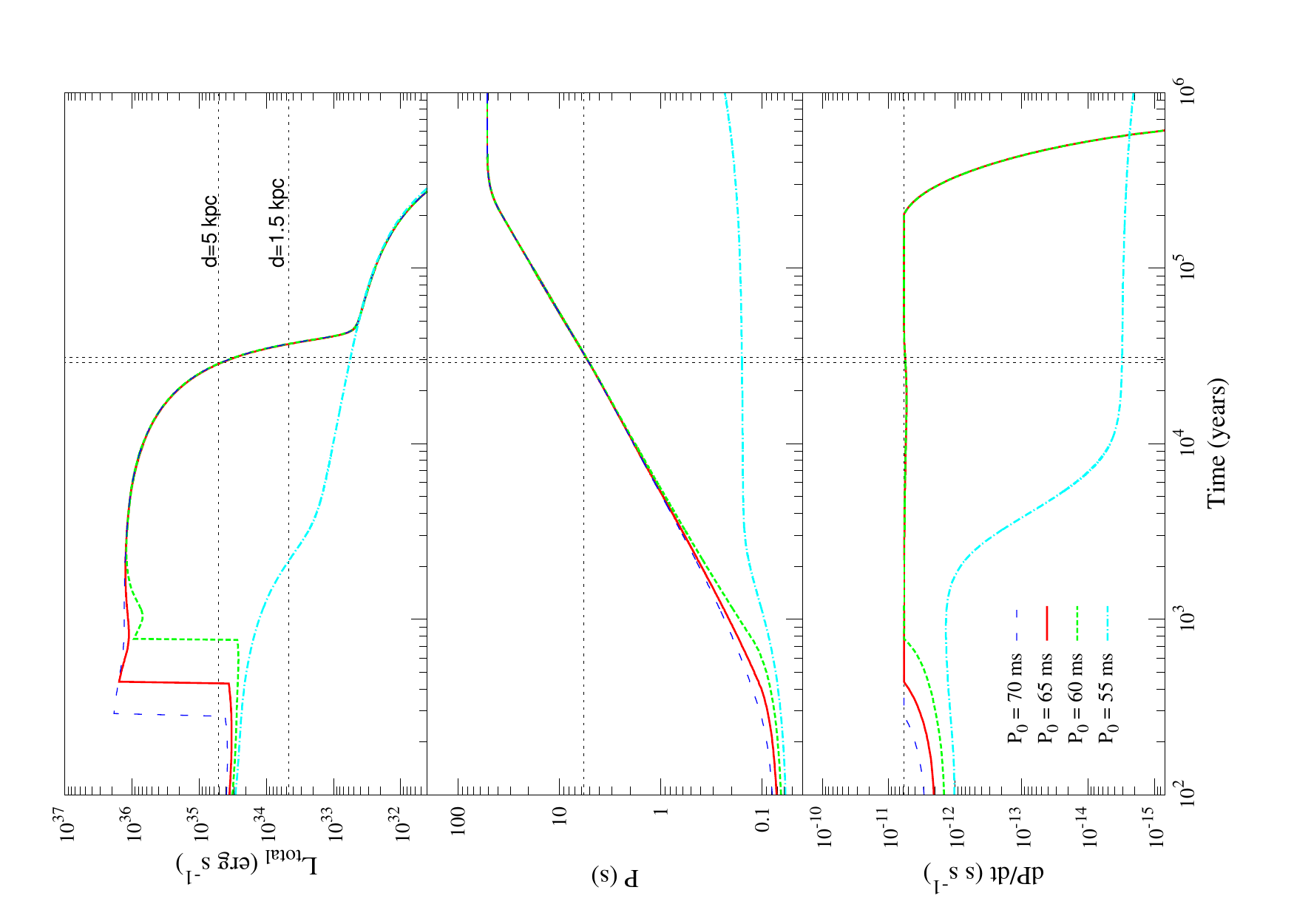}
\caption{Model curves with the same parameters as those of the dashed curve in Fig. 1, but with different initial periods ($P_0$) given in the bottom panel. It is seen that the model results are not sensitive to $P_0$ if $P_0\mathrm{\gtrsim}60$ ms. For lower values of $P_0$, the sources cannot enter the light cylinder and are likely to evolve as radio pulsars that cannot reach the properties of \0501.} 
\end{figure}

In the simulations, we take the initial period $P_0 = 300$ ms. The long-term evolution of a model source is not sensitive to $P_0$ provided that the star enters the long-term accretion phase. The model light curves shown in Fig. 2 illustrate evolutions with different $P_0$ values. We repeat the simulations tracing the $P_0$ values to find the minimum $P_0$ that allows the onset of the accretion. We find that the minimum $P_0$ for \0501~is $\sim 60$ ms (see Fig. 2).

\begin{figure}
\includegraphics[height=.24\textheight,angle=0]{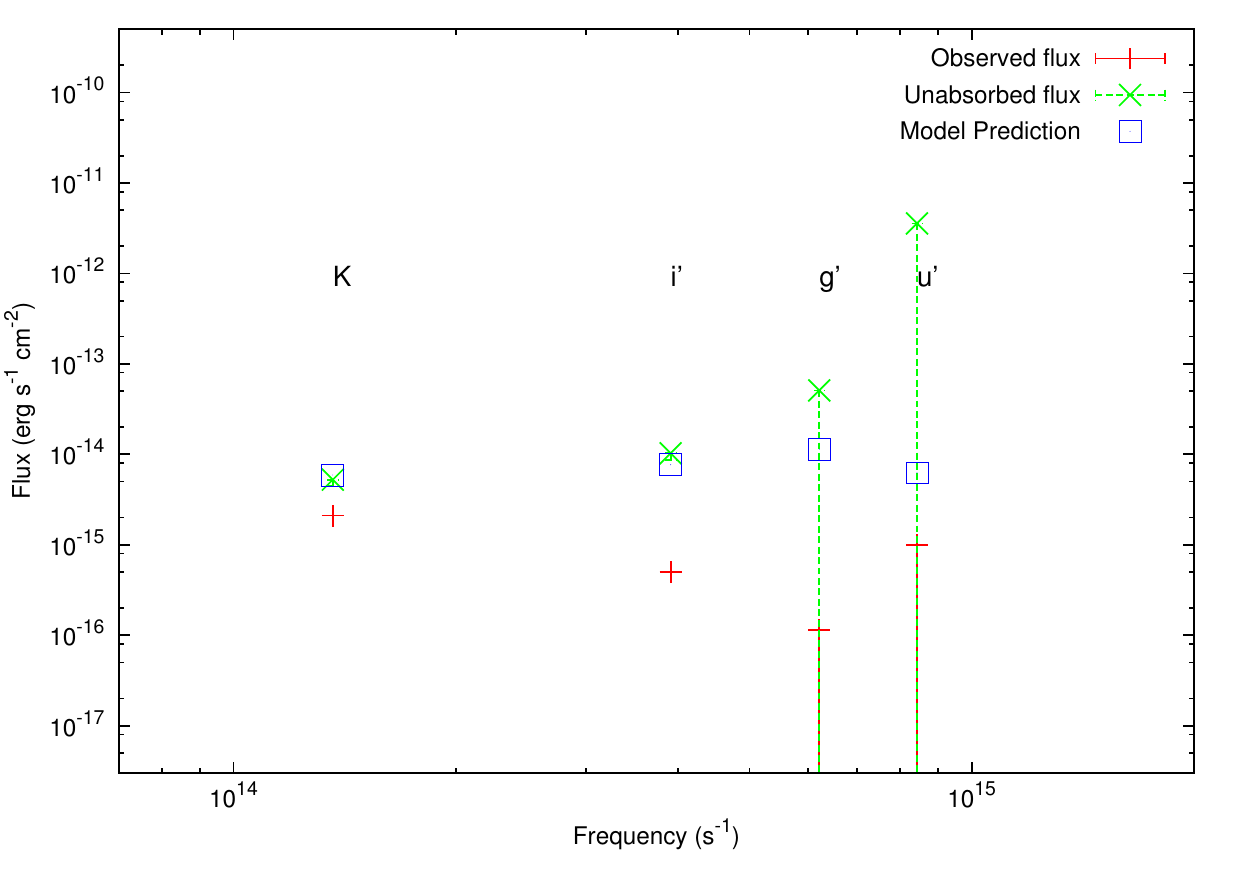}
\caption{Optical/IR emmission of \0501. The blue boxes show the model predictions of an irradiated fallback disc in four different energy bands. $\Mdotin = 1.4 \times 10^{14}$ g s$^{-1}$, disc inclination angle is $\sim 70\,^{\circ}$ and irradiation efficiency $C = 10^{-4}$. For the bands $g'$ and $u'$ the data show $3\sigma$ upper limits. } 
\end{figure}

For \0501 assuming that the inner disc extends down to the co-rotation radius, we estimate $i'$, $u'$, $g'$ and $K$ band fluxes for a distance of 1.5 kpc using the model described in Ertan \& \Caliskan~(2006). For $N_\mathrm{H} = 10^{22}$ cm$^{-2}$ (Rea et al. 2009) we have converted the observed magnitudes when the X-ray luminosity is close to the quiescent level [$i'=24.4\pm0.4$, $u'>24.7$, $g'>26.9$, $K=19.7\pm0.1$ (Dhillon et al. 2011)] into the unabsorbed flux values. It is seen in Fig. 3 that the irradiated fallback disc model is in good agreement with optical/IR data. We obtain the model fluxes given in Fig. 3 with $C = 1 \times 10^{-4}$ and $\Mdot = 1.4 \times 10^{14}$ g s$^{-1}$ which are consistent with long-term evolution and the short-term X-ray enhancement models within the distance uncertainties (see Fig. 1).

\section{X-ray Outburst of \0501}

The details and applications of our X-ray enhancement model can be found in \Caliskan~\& Ertan (2012). The model can be summarized as follows: The disc in the quiescent state mimics a steady state geometrically thin disc and evolves with $\rin = \rA$. Part of the energy emitted during a burst episode moves the inner disc matter to a larger radii. This causes the inner disc mater to pile-up at the inner disc which we represent by a Gaussian mass distribution as the initial condition in our model. The evolution of this density gradient first leads to an abrupt increase in the mass-flow rate of the inner disc, the accretion rate on to the star and thus the X-ray luminosity. Subsequently, the X-ray luminosity  decreases with a rate governed by the viscous relaxation of the disc yielding the decay phase of the X-ray luminosity. 

Early decay phase ($\sim 150$ d) of the X-ray outburst of \0501 was investigated earlier by  \Caliskan~\& Ertan (2012).  With the addition of the new data, covering the time period from $\sim 150$--$600$ d after the burst (Camero et al. 2014), we revisited the model fits, to test the model predictions from the maximum to the end of the decay phase. The available absorbed data for the first $\sim 150$ d of the decay phase were obtained from different satellites, and with fits to different spectral models. Due to difficulties in obtaining the unabsorbed X-ray data of this source in a systematic way, \Caliskan~\& Ertan (2012) used the absorbed X-ray data considering also that the absorption does not significantly affect the light curve close to the peak of the outburst when the X-ray luminosity and the temperatures are sufficiently high.  

From 150 to 600 d of the decay phase, the X-ray light curve gradually decreased to the quiescent X-ray flux level of the source (Camero et al. 2014). In this late decay phase, increasing absorption with decreasing luminosity probably alters the light curve morphology significantly. In the present work, we use only the six XMM-Newton unabsorbed X-ray data points of \0501~covering the whole decay phase. A similar study was also done by Camero et al (2014, fig. 11) applied to the outburst model of Pons \& Rea (2012). The first five data points are taken from Rea et al. (2009) and the sixth data point is taken from Camero et al (2014). 

Figs. 4 and 5 show the $0.5$--$10$ keV unabsorbed flux of \0501~obtained from the six XMM-Newton observations of the source together with our model fits for the distances 1.5 and 5 kpc respectively. For a given distance, we give model curves for two different irradiation efficiencies (red solid and green dashed). It is seen in Figs. 4 and 5 that the model light curves are in very good agreement with the X-ray data. 

\begin{figure}
\includegraphics[height=.35\textheight,angle=270]{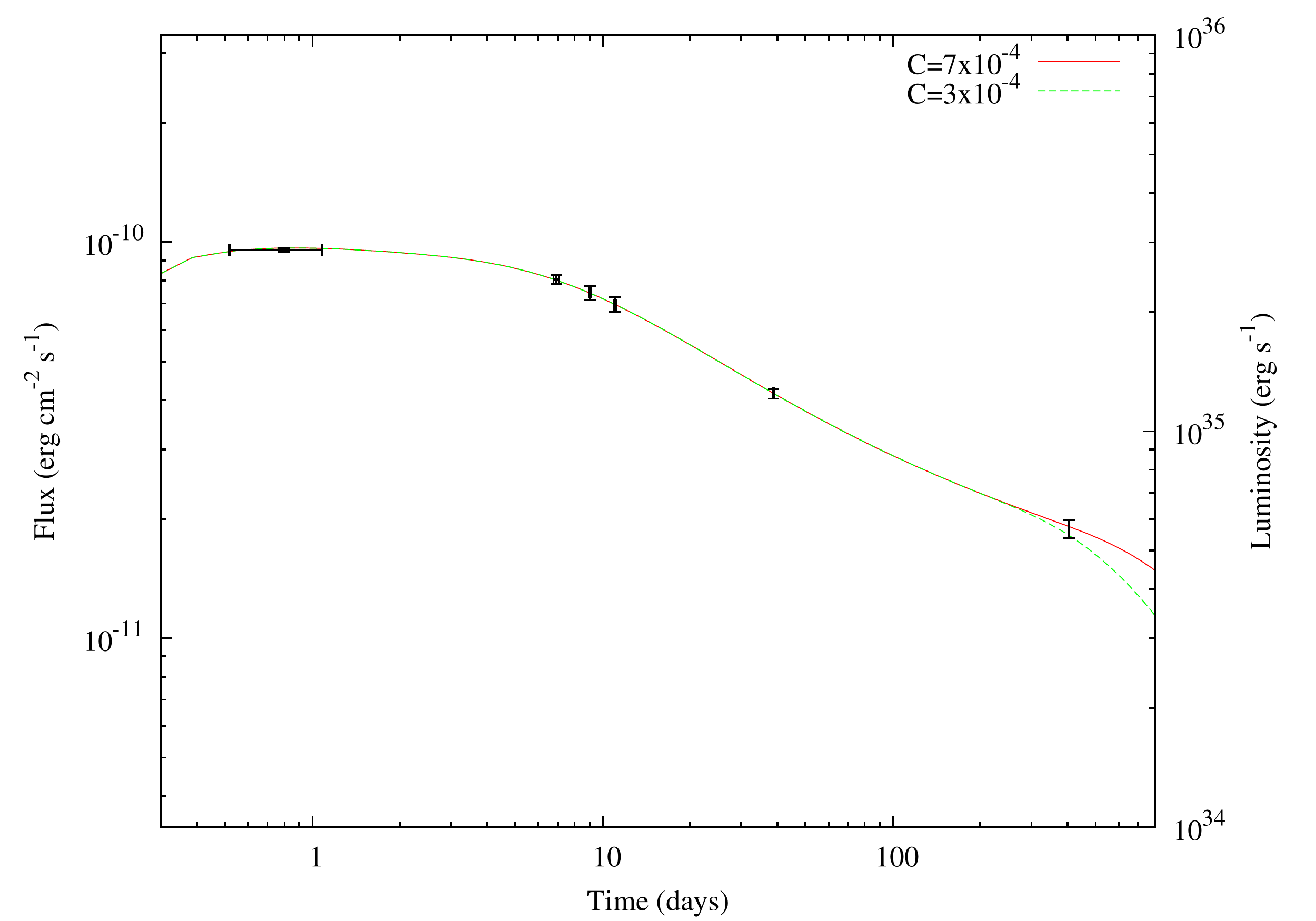}
\caption{X-ray outburst decay lightcurve of \0501. The unabsorbed 0.5--10 keV flux data points are from XMM-Newton observations. The first five data points are taken from Rea et al. (2009) and the sixth data point is taken from Camero et al (2014). The X-ray luminosity is calculated assuming a distance of 5 kpc. The horizontal error bar of the first data point denotes the time interval of observation.} 
\end{figure}

\begin{figure}
\includegraphics[height=.35\textheight,angle=270]{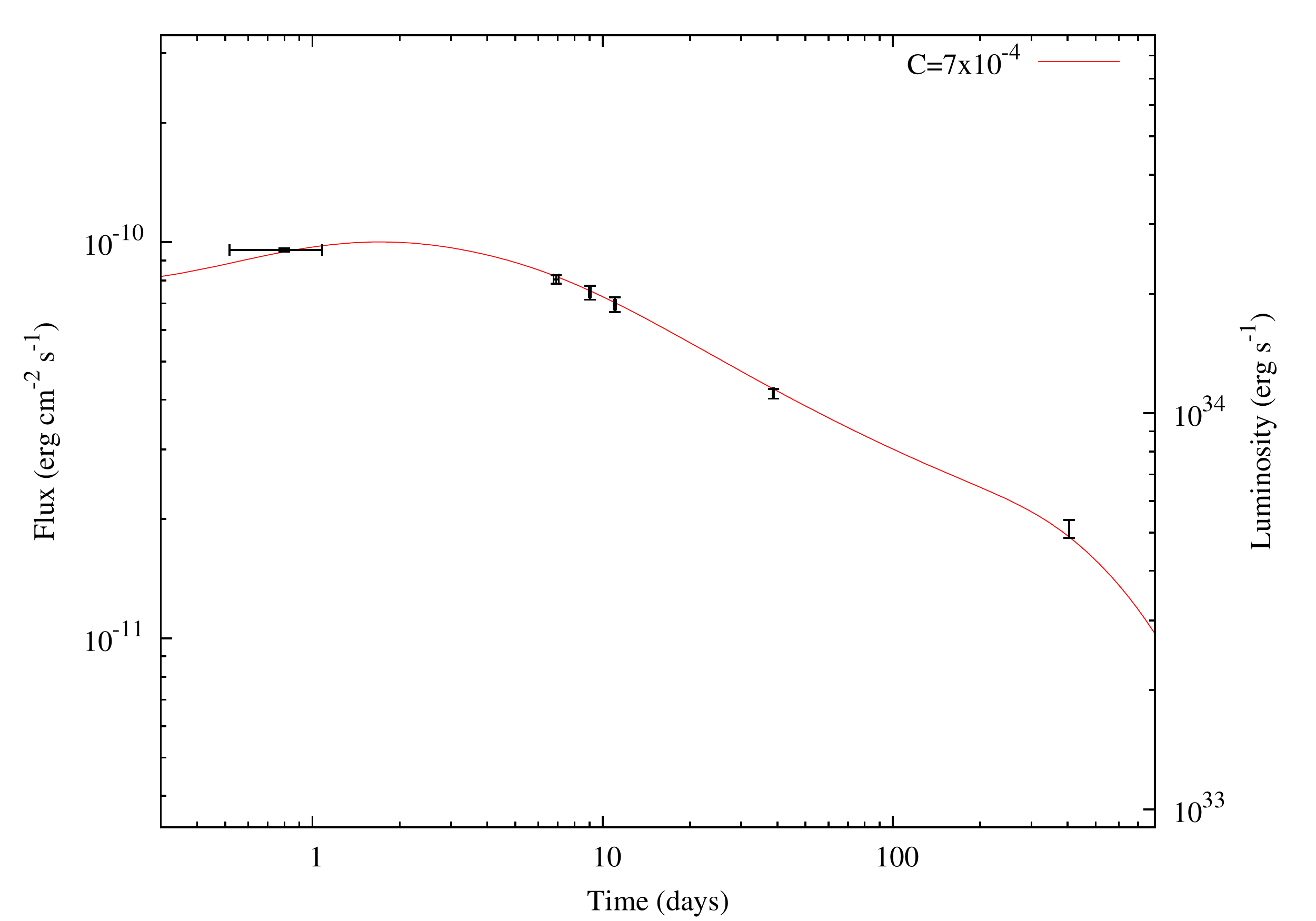}
\caption{The same as Fig. 4, but for a source distance of 1.5 kpc. This model curve is obtained with the irradiation efficiency $C = 7 \times 10^{-4}$. Smaller irradiation efficiencies do not give reasonable fits for this particular distance.} 
\end{figure}

\begin{table} \caption{\label{table:params}The parameters for the models
presented in Fig. 4 and 5. The basic disc parameters  $\ah = 0.1$, $\ac = 0.045$, $\Tcrit = 1750$ K are the same for the model curves given in Figs. 4 and 5. The irradiation efficiency $C = 7 \times 10^{-4}$ for $d = 1.5$ kpc and $C =  3$--$7 \times 10^{-4}$ for $d = 5$ kpc give good fits to data. See \Caliskan~\&  Ertan (2012) for a detailed explanation of the model parameters.  
}

\begin{minipage}{\linewidth} 
\begin{center} 
\begin{tabular}{c|c|c} \hline \hline 
Parameter & $d = 1.5$ kpc & $d = 5$ kpc\\ \hline 
$\rin$~(cm) & $3.1 \times 10^{9}$ & $3 \times 10^{9}$\\ 
$\rzero$~(cm) & $6 \times 10^{9}$ & $6 \times 10^{9}$\\ 
\dr (cm) & $7 \times 10^{8}$ & $3 \times 10^{9}$\\
$\Smax$ (g cm$^{-2}$) & 4.1 & 14.4\\ 
$\Szero$ (g cm$^{-2}$) & 2.7 & 13.1\\ 
$C$ & $7 \times 10^{-4}$ & 3--$7 \times 10^{-4}$\\ 
\hline \hline 
\end{tabular}\\

\end{center} \end{minipage} \end{table}

We note that the viscosity parameters $\ac$ and $\ah$, irradiation strength $C$ and the critical temperature $\Tcrit$ are the basic disc parameters that are expected to be the same for different AXP/SGRs in the same accretion regime within the distance uncertainties. For comparison with the results obtained by \Caliskan~\& Ertan (2012), we present our model parameters in Table 1.

\section{CONCLUSIONS} 

We have investigated the possible long-term evolutionary scenarios and the short-term X-ray outburst light curve of \0501 in the fallback disc model. Our results show that a neutron star with a fallback disc and with a dipole field strength of $\sim 1.4 \times 10^{12}$ G on the pole of the star acquires the X-ray and the rotational properties of \0501 in $\sim 3 \times 10^4$ yr. We have also shown that the X-ray enhancement light curve of the source can be reproduced by the evolution of the inner disc after the soft gamma burst episode. 

For both the long-term evolution and the X-ray enhancement models, the source properties are achieved with the same basic parameters used earlier to explain the enhancement curves of other AXP/SGRs (\Caliskan~\& Ertan 2012) and the long-term evolution of \ql high-B radio pulsar\qr~PSR J1734--3333, the low-B magnetars and the dim isolated neutron stars (Ertan et al. 2014). This indicates that in the fallback disc model, all these apparently rather different sources are actually the neutron stars with fallback discs and conventional dipole fields in different phases of their long-term evolution. The model is also consistent with the radio properties of these sources. For PSR J1734--3333, our results imply that the source is evolving in the early phase before the onset of accretion with a rotational rate sufficiently high for radio emission. The low-B magnetars, like dim isolated neutron stars (Ertan et al. 2014), are evolving in the tracking phase after the  accretion epoch, and located below the pulsar death line on the $P$--$B$ plane.   

For \0501, we do not expect pulsed radio emission, since the source is currently accreting matter from the fallback disc. Our results indicate that this source will find itself below the pulsar death line at the end of its accretion phase. Even at present, if accretion is hindered by any reason, the rotational rate of the source with $B_0 \sim 1.4 \times 10^{12}$ G is not sufficient to produce pulsed radio emission. 

Furthermore, we have shown that the observed optical/IR properties of \0501 are in agreement with the spectrum expected from a fallback disc with current properties indicated by our long-term evolutionary model.

\section*{Acknowledgments}

We acknowledge research support from Sabanc\i\ University, and from
T\"{U}B{\.I}TAK (The Scientific and Technological Research Council of
Turkey) through grant 113F166.

\bsp

\label{lastpage}

\end{document}